\documentclass[aps,preprint,english]{revtex4-1}
\usepackage[latin9]{inputenc}
\usepackage{float}
\usepackage{textcomp}
\usepackage{amsmath}
\usepackage{amssymb}
\usepackage{graphicx}

\makeatletter

\@ifundefined{textcolor}{}
{%
 \definecolor{BLACK}{gray}{0}
 \definecolor{WHITE}{gray}{1}
 \definecolor{RED}{rgb}{1,0,0}
 \definecolor{GREEN}{rgb}{0,1,0}
 \definecolor{BLUE}{rgb}{0,0,1}
 \definecolor{CYAN}{cmyk}{1,0,0,0}
 \definecolor{MAGENTA}{cmyk}{0,1,0,0}
 \definecolor{YELLOW}{cmyk}{0,0,1,0}
}

\makeatother

\usepackage{babel}
\begin{document}

\title{Two-Dimensional hybrid perovskites sustaining strong polariton interactions
at room temperature}

\author{A. Fieramosca$^{1,2}$\footnote{These autors equally contributed to this work},
L. Polimeno$^{1,2,4}$\footnotemark[1], V. Ardizzone$^{1,2}$\footnote{Electronic Address: v.ardizzone85@gmail.com},
L. De Marco$^{1,}$\footnote{Electronic Address: luisa.demarco@nanotec.cnr.it},
M. Pugliese$^{1}$, V. Maiorano$^{1}$, M. De Giorgi$^{1}$, L. Dominici$^{1}$,
G. Gigli$^{1,2}$, D. Gerace$^{3,1}$, D. Ballarini$^{1}$, D. Sanvitto$^{1,4}$.}

\affiliation{$^{1}$CNR Nanotec, Institute of Nanotechnology, via Monteroni, 73100
Lecce, Italy.}

\affiliation{$^{2}$Dipartimento di Matematica e Fisica, Universit\'{a} del Salento,
via Arnesano, 73100 Lecce, Italy.}

\affiliation{$^{3}$Dipartimento di Fisica, Universit\'{a} degli Studi di Pavia, via
Bassi 6, 27100 Pavia, Italy.}

\affiliation{$^{4}$INFN Istituto Nazionale di Fisica Nucleare, Sezione di Lecce,
73100 Lecce, Italy.}
\begin{abstract}
Polaritonic devices exploit the coherent coupling between excitonic
and photonic degrees of freedom to perform highly nonlinear operations
with low input powers. Most of the current results exploit excitons
in epitaxially grown quantum wells and require low temperature operation,
while viable alternatives have yet to be found at room temperature.
Here we show that large single-crystal flakes of two-dimensional layered
perovskite are able to sustain strong polariton nonlinearities at
room temperature with no need to be embedded in an optical cavity.
In particular, exciton-exciton interaction energies are measured to
be remarkably similar to the ones known for inorganic quantum wells
at cryogenic temperatures, and more than one order of magnitude larger
than alternative room temperature polariton devices reported so far.
Thanks to their easy fabrication, large dipolar oscillator strengths
and strong nonlinearities, these materials hold great promises to
realize actual polariton devices at room temperature. 
\end{abstract}
\maketitle

\section{introduction}

In efficient communication and computing systems, information carriers
are required to both travel long distances without losing coherence
and simultaneously interact between them in order to implement logic
functions such as switches or logic gates. Microcavity polaritons,
quasiparticles that form in a semiconductor when an elementary excitation
field interacts sufficiently strongly with the electromagnetic radiation
field, could in principle fulfill these requirements and are promising
candidates for a new generation of optoelectronic devices. Indeed,
these \textquotedblleft dressed\textquotedblright{} photons exhibit
enhanced non-linearities thanks to their electronic component \citep{PhysRevLett.84.1547,PhysRevLett.69.3314},
allowing many-body effects to be studied in optical systems and appearing
as building blocks for integrated photonic circuits and electro-optic
applications \citep{RevModPhys.85.299,Sanvitto2016}. However, to
bring strong nonlinearities and good optical properties to room temperature
is a real challenge. Recently, polariton condensation and superfluidity
have been observed at room temperature (RT) in organic semiconductors
thanks to the very large binding energy and oscillator strength of
Frenkel excitons \citep{Lerario2017}. However, interactions among
excitons in organic materials are usually at least two orders of magnitude
lower than the ones observed for typical Wannier-Mott excitons, and
the spin degree of freedom, that is of paramount importance for photonic
applications, is quickly averaged out. On the other hand, polaritons
in layered transition metal dichalcogenide (TMD) materials have been
anticipated to be a possible way-out, as recently demonstrated \citep{Barachati2018},
showing interactions at least ten times higher than organic materials.
Howewer, they require to work with single monolayers, difficult to
control and hardily reaching extensions of more than several tens
of \textmu m. 

From a wider perspective, hybrid organic-inorganic perovskite systems
have recently attracted a considerable attention driven by exceptional
progress in photovoltaics, photonics, and optoelectronics \citep{doi:10.1021/acsnano.6b05944,Saparov2016,Sutherland2016,Thouin2018}.
In particular, two-dimensional (2D) perovskites are spontaneous realizations
of multiple layered quantum well (QW) hetero-structures made of $[PbX_{6}]2-$
tetrahedral inorganic layers, with X indicating an halide, sandwiched
between bilayers of organic cations (see figure 1a) \citep{Niu2014,Yaffe2015}.
The lowest-energy electronic excitations are associated to the inorganic
sheet, the organic part acting as a potential barrier \citep{doi:10.1021/acsnano.6b05944,Saparov2016,PhysRevB.42.11099,Kondo1998}.
In fact, these 2D layered structures share strong similarities with
multi-quantum well heterostructures made of epitaxially grown inorganic
semiconductors, but they possess larger binding energies and display
stronger dielectric confinement in the inorganic layers. 2D hybrid
perovskites show also enhanced collective effects due to the large
number of layers stacked in a single crystal, which is an advantage
over TMD monolayers. Strong light-matter coupling has already been
observed in $MAPbBr_{3}$ micro/nanowire cavities \citep{Zhang2017};
in particular, evidence for exciton-polariton effects has been reported
in all-inorganic perovskite $CsPbX_{3}$ nanoplatelets, nanowires
\citep{doi:10.1021/acs.nanolett.7b01956,doi:10.1021/acs.jpclett.6b01821}
and in spin-coated layered hybrid organic-inorganic perovskite thin
films \citep{PhysRevB.57.12428,Pradeesh:09,Brehier2006}. These results
indicate a huge room for substantial advancements in the realization
of low threshold coherent light sources and polaritonic devices with
this kind of hybrid organic-inorganic semiconductors \citep{Low2016}. 

Here we address, for the first time, the nonlinear properties of polaritons
in monocrystalline 2D hybrid perovskites by directly estimating the
exciton-exciton interaction energy. The understanding and control
of this property is of paramount importance to assess the material
potentialities for room temperature polaritonics, such as all-optical
transistors, switches and gates, up to now only demonstrated at cryogenic
temperatures in GaAs based semiconductors. To this aim, we employ
large flakes of high-quality 2D perovskite single crystals to directly
measure the excitonic response without being limited by non-radiative
losses and grain-to-grain heterogeneity, usually present in polycrystalline
films. Remarkably, at room temperature we find the exciton oscillator
strengths to be much larger than in epitaxial GaAs QWs at cryogenic
temperatures, and with comparable nonlinearities. Most importantly,
thanks to the large oscillator strength of the excitons, and the very
good crystal structure of the flakes, formation of polaritonic bands
and strong nonlinearities are observed even when these structures
are not embedded in a microcavity. This is an important step towards
the realization of polaritonic devices working at room temperature
with easy fabrication and minimal processing, avoiding complex heterostructures
comprising top and bottom Bragg mirrors. 

\section{Results and Discussion}

We synthesize phenethylammonium lead iodide $(C_{6}H_{5}(CH_{2})_{2}NH_{3})_{2}PbI_{4}$
(PEAI) by an Anti-solvent Vapor assisted Crystallization method \citep{Ledee2017}
and subsequent mechanical exfoliation \citep{Niu2014}. In this work,
we study both PEAI single crystals embedded in an optical cavity formed
by two Distributed Bragg Reflectors (DBR) acting as bottom/top mirrors
and bare single crystals. Figure 1a shows a sketch of a single 2D
PEAI crystal grown on a DBR formed by seven $Si0_{2}/Ti0_{2}$ pairs.
A second DBR also formed by seven $Si0_{2}/Ti0_{2}$ pairs is sputtered
on the single crystal in order to give rise to an optical Fabry-Perot
cavity in which the PEAI 2D crystal is embedded. The optical response
of the sample is measured in transmission geometry, the excitation
being provided by a non-resonant continous wave laser or a femtosecond
pulsed laser resonant on the low polariton branch (see also Supplementary
Information). 

\begin{figure}[H]
\centering{}\includegraphics[scale=0.4]{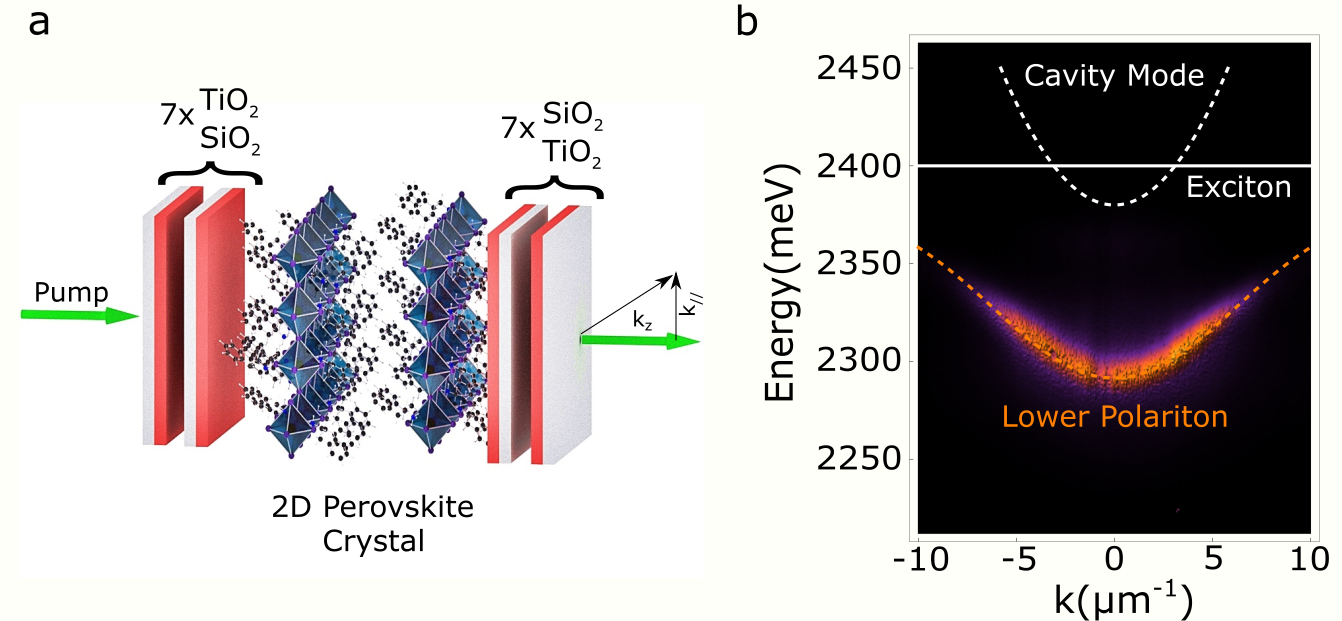}\caption{a) Schematic representation of a 2D perovskite single crystal embedded
in an optical cavity formed by two distributed Bragg reflectors; in
2D perovskite inorganic layers are separated by organic ligands realizing
an effective multiple layered quantum well structure; b) Energy versus
in-plane momentum $k$ photoluminescence emission from from the sample
represented in figure 1a, with $k=\frac{2\pi}{\lambda}sin\theta$,
$\theta$ being the emission angle; the white dashed and solid lines
represent respectively the cavity and the exciton uncoupled modes,
the orange line is a fit to the polariton lower mode with $E_{X}=2.395$
eV, $E_{C}=2.385$ eV and $\hbar\Omega=170$ meV;}
\end{figure}

Figure 1b shows the energy and in-plane momentum resolved emission
from this sample under non-resonant excitation. Due to the strong
coupling between the cavity mode and the exciton in the PEAI single
crystal, a lower polariton mode is clearly visible with an emission
energy of about 2.31 eV at $k$ = 0. By fitting the experimental dispersion
we obtain a Rabi splitting $\hbar\varOmega=170$ meV, with an energy
minimum of the cavity mode at $E_{C}=2.385$ eV and an exciton energy
$E_{X}=2.395$ eV. 

To test the polarization dependent polariton-polariton interaction
in this sample, we measure the resonant transmission of a fs pulsed
excitation laser through the sample at normal incidence (k=0). Figure
2a shows the transmission when the laser is linearly polarized and
figure 2b shows the transmission when the laser is circularly polarized.
Each spectrum in figure 2a and 2b corresponds to different excitation
powers, $P$. As the excitation power increases, we observe a shift
of the transmission peak towards higher energies. The energy blueshift
of the transmission peaks is plotted in figure 2c against the excitation
power. As the excitation power increases, we observe a shift of the
transmission peak towards higher energies as expected for a system
of interacting particles \citep{PhysRevLett.99.140402,doi:10.1021/acs.nanolett.7b01956},
the polariton density inside the active medium being proportional
to the incident power. Moreover, figure 2c shows that the energy blueshift
obtained with a circularly polarized laser (blue dots) is higher than
the blueshift measured with a linearly polarized laser (red dots).
In other words, the observed energy blueshift is sensitively larger
when all the polaritons are created with the same spin \citep{PhysRevB.62.R4825,PhysRevB.82.075301,PhysRevB.79.115325,Ballarini2007}.
Instead, by using a linear polarized laser (i.e. a coherent superposition
of two counter-polarized circular components), only half of the excited
polaritons share the same spin, thus roughly halving the interaction
energy. We notice that the polarization properties of the elementary
excitations in these materials arise from the optical transitions
connecting the valence band s-type states with total angular momentum
quantum numbers ($\ensuremath{J=1/2},\ensuremath{J_{z}=\pm1/2}$)
to p-type conduction band states with the same symmetry (due to the
spin-orbit splitting in the conduction band) \citep{Fieramosca2018,Giovanni2016}.
Hence, bright exciton states correspond to optical transitions satisfying
the condition $\ensuremath{\Delta J_{z}=\pm1}$, which can be excited
by using either clockwise or counter-clockwise circularly polarized
radiation. As a consequence, polariton states are formed with $\pm1$
spin polarization, in close analogy to inorganic semiconductor exciton-polaritons
in microcavities with embedded quantum wells. 

The observed spin-dependence is then a strong indication that polariton-polariton
interactions not only are the dominant effect responsible for the
energy shift of the modes upon optical excitation but also that 2D
perovskites can be used for polariton spintronics. The dashed lines
in figure 2c are obtained as a linear fit to the experimental points.
The ratio between the two slopes is $L/C=0.47\pm0.06$. By following
Vladimirova et al. \citep{PhysRevB.82.075301} we can extract the
ratio $\alpha_{2}/\alpha_{1}$ from the data of figure 2c, in which
$\alpha_{2}$ is the interaction strength between polaritons having
opposite spin and $\alpha_{1}$ is the interaction strength between
polaritons having the same spin. We deduce then from figure 2c a ratio
$\ensuremath{|}\alpha_{2}/\alpha_{1}|\sim0.05.$ This value is consistent
with the picture of polariton-polariton interactions in standard semiconductors
at cryogenic temperature, i.e., a strong repulsive interaction for
polaritons having the same spin ($\alpha_{1}>0$) and a weaker interaction
for polaritons having opposite spin ($\ensuremath{|}\alpha_{1}\ensuremath{|}\gg\ensuremath{|}\alpha_{2}\ensuremath{|}$).
Strikingly, we observe spin-dependent nonlinearities similar to those
observed in GaAs based systems at cryogenic temperatures. 

In the mean-field approximation and at low particle density the blueshift
$\triangle E_{pol}$ depends linearly on the polariton density $n_{pol}$:
$\triangle E_{pol}=g_{pol}n_{pol}$. The polariton density $n_{pol}$
is proportional to the excitation power incident on the sample. By
analyzing the power dependence of the blueshift it is possible to
measure the polariton interaction constant and then, knowing the polariton
excitonic fraction (or Hopfield coefficient) one can finally infer
the excitonic interaction constant $g_{exc,lay}$ per layer. The estimated
excitonic interaction constant values for the microcavity sample is
$g_{exc,lay}\sim3\pm0.5$ $\mu eV\mu m^{2}$ (Supplementary Information).
This value is comparable with the one typically estimated for a single
GaAs quantum well \citep{Walker2015,PhysRevLett.119.097403} and it
is the largest reported value at room temperature so far. 

\begin{figure}[H]
\centering{}\includegraphics[scale=0.3]{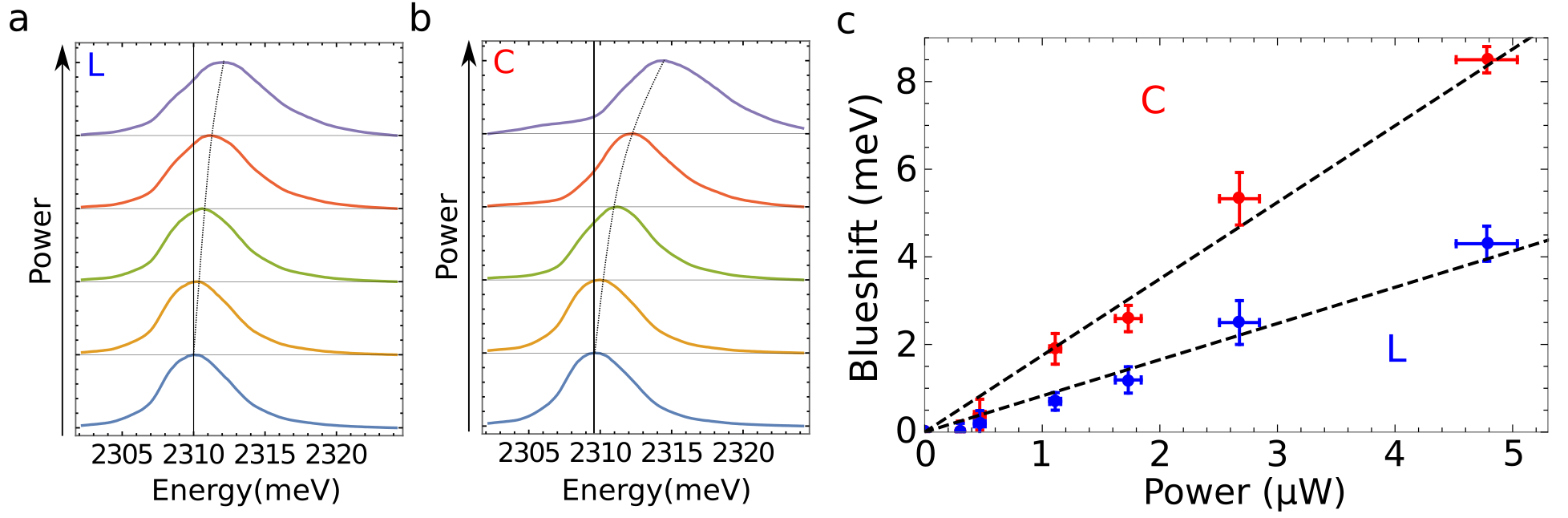}\caption{a) Transmittivity spectra obtained by cutting the dispersion of figure
1b in $k=0$ and corresponding to different resonant excitation power
for linear (a) and circular (b) polarized excitation laser; c) Blueshift
of the polariton modes in the case of a linear (L) and a circular
(C) polarized laser; the dashed lines are linear fit to the experimental
data with slopes of 1.75 and 0.83 $meV/\mu W$ for C and L respectively.}
\end{figure}

The observation of polaritonic nonlinearities at RT in organic-inorganic
hybrid materials is a fundamental step to assess the potential for
real world polariton devices. The structure of the sample of figure
1a is complex and requires at least a three-steps fabrication process:
growth of the bottom DBR mirror, growth of the single crystal flake,
and finally growth of the top DBR mirror. Further developments could
involve additional fabrication steps like electrical contacts for
carriers injection or interconnecting several devices to provide the
different functionalities associated with network-on-chip technology.
These further steps would result in an even more complex fabrication
process, possibly hindering the technological appeal of RT polaritonic
devices. However, RT polaritons in 2D single crystal PEAI flakes have
been observed even if the crystal flake is not embedded in an optical
cavity \citep{Fieramosca2018}. Single bare PEAI crystal flakes without
cavity are a relatively simpler system to grow. Moreover, the active
region can be readily accessed for electrical connections, patterning
or for interconnecting several devices. From a technological point
of view, assessing RT polariton nonlinearities in single 2D crystals
with no cavity, is of utmost importance and would lift the complication
to embed the perovskite between two mirrors.

Figure 3a shows a sketch of a single 2D PEAI crystal flake grown on
a glass substrate. Figure 3b shows in-plane momentum and energy resolved
reflectivity spectra obtained from the single 2D PEAI crystal slab
under white light illumination. The optical response of the sample
is measured with an oil-immersion microscope objective which allows
to capture, from the glass substrate side, also the signal from the
total internal reflection (TIR) at the air-crystal interface (see
Supplementary Information for details). Remarkably, the measured reflectivity
shows a manifold of lower polariton modes (visible as dips in the
reflectivity) arising from the coupling of the bare excitonic resonance
(white line) with the optical resonances of the 2D slab itself \citep{Fieramosca2018}.
The red dashed lines of figure 3b are fits to the lower polariton
modes giving a value of energy coupling between the optical modes
and the excitonic transition of $\ensuremath{\hbar\Omega}\sim170$
meV. This value is larger than the linewidths of the different resonances
involved and it is then fully consistent with the strong exciton-photon
coupling regime \citep{Fieramosca2018}. In particular, strong coupling
between the exciton mode and the optical resonances of the slab is
obtained at room temperature thanks to the optical confinement given
by the refractive index contrast at the interface between the crystal
and the surrounding media (respectively air and glass substrate) and
the high value of the excitonic oscillator strength in 2D perovskites. 

\begin{figure}[H]
\begin{centering}
\includegraphics[scale=0.45]{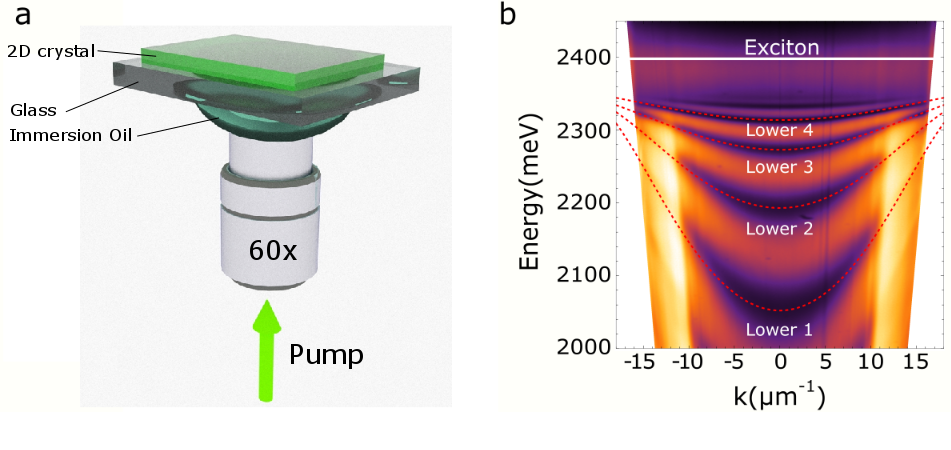}
\par\end{centering}
\caption{a) A schematic representation of the TIR configuration adopted for
resonant blueshift measurements: an immersion oil objective (60x)
is used to focus the excitation beam on a 2D PEAI single crystal flake
grown on a glass substrate; the same objective is used to collect
the reflected light; b) Energy and in-plane momentum $k$ resolved
reflectivity spectra of a thick single crystal slab of PEAI; the dips
in reflectivity correspond to lower polariton modes resulting from
the coupling of the exciton mode to different optical modes; the white
line represents the energy of the bare exciton mode; the red dashed
lines represent lower polariton modes; the enhanced intensity for
$k\geqslant10$ $\mu m^{-1}$ corresponds to angles of incidence beyond
the light line between air and the perovskite slab. }
\end{figure}

In order to understand if the observed polariton modes of the bare
2D PEAI crystal can sustain nonlinearities at RT, we resonantly excite
the single crystal slab with a pulsed fs laser. The excitation beam
forms a finite angle with the surface of the slab, corresponding to
an in-plane momentum of about $k\sim12.7$ $\mu m^{-1}$. This value
of in-plane momentum corresponds to an angle that is beyond the crystal-air
TIR angle, which means that all the incident laser power is absorbed
or reflected by the sample, the transmitted part being negligible.
The green oval region in figure 4a shows the energy and momentum spread
of the laser used in resonant excitation. Figure 4b represents a vertical
cut of the dispersion of figure 4a corresponding to the green region. 

This choice of the incident angle, in addition to being in the highest
reflectivity region, allows also to probe polariton modes having a
higher excitonic fraction and smaller linewidth. The bandwidth of
the fs laser is large enough to resonantly excite essentially three
adjacent lower polariton branches. Figure 4b shows that the reflectivity
spectra change when the incident power is increased: the low-power
resonances (red continuous lines) are shifted at high incident power
(blue continuous lines) and then recover their initial energy (red
dashed lines) when the incident power is lowered\textendash which
guarantees that the blueshift is not due to degradation of the material
upon optical excitation. This evidence shows that the observed blueshift
arises from interparticle interactions, as observed in low-temperature
inorganic polariton systems \citep{PhysRevB.79.115325,PhysRevB.82.075301,PhysRevB.62.R4825}
and in the cavity-embedded single crystal of figure 2.

\begin{figure}[H]
\begin{centering}
\includegraphics[scale=0.2]{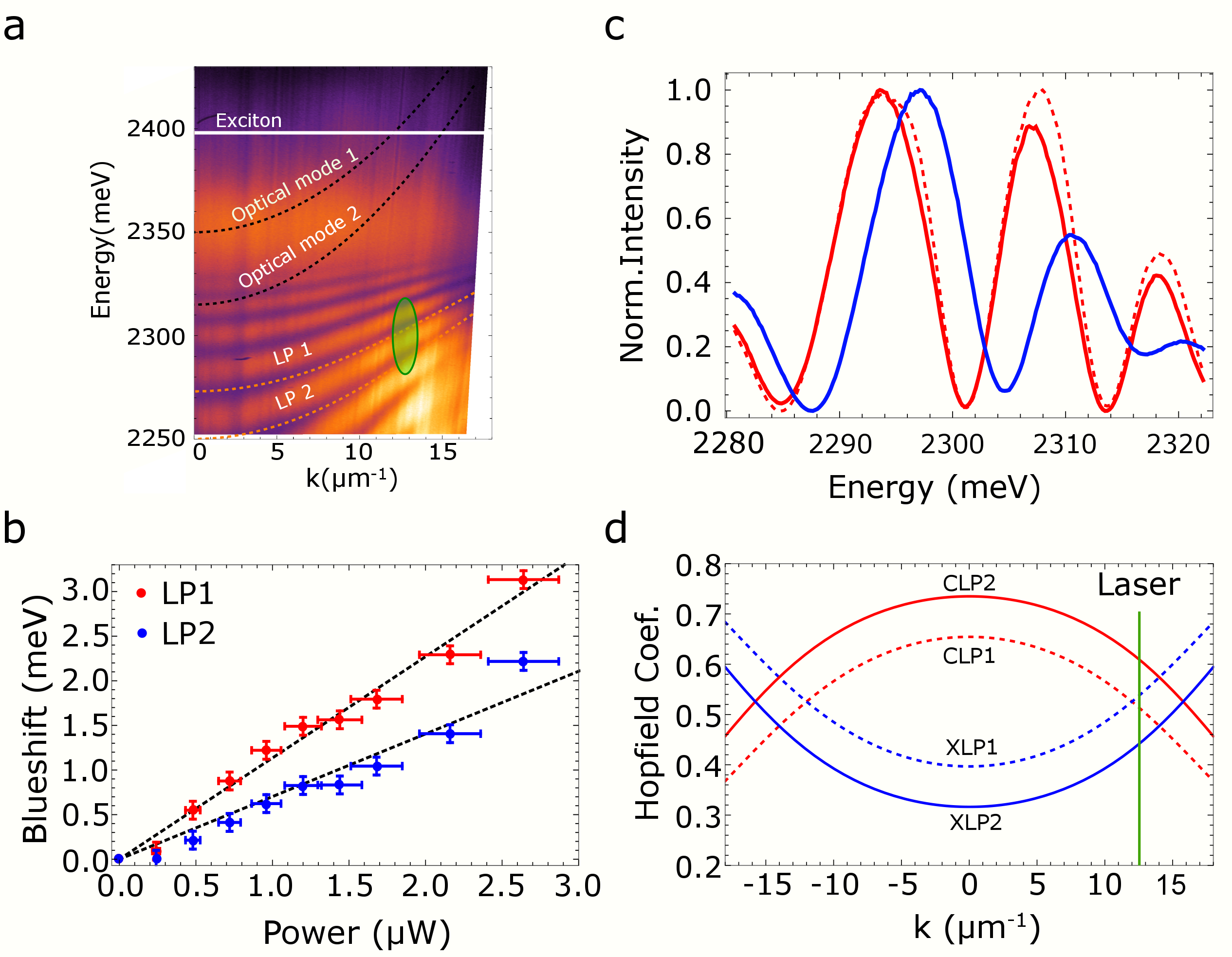}
\par\end{centering}
\caption{a) Reflectivity spectra obtained for a PEAI single crystal slab; the
same excitonic mode is coupled to several optical modes of the slab;
the two lower polariton modes highlighted by orange dashed lines originate
from the coupling of the two optical modes (black dashed lines) to
the excitonic mode (white solid line); b) energy shift of the lower
polariton modes of (a) as a function of the incident power and; the
energy shift is measured at $k\sim12.7$ $\mu m^{-1}$ corresponding
to the green region of panel (a); c) reversibility of the observed
blueshift of the polariton modes; the red continuous curve correspond
to the energy of the polariton modes for low excitation power $P=10$
$\mu J/cm^{2}$; the continuous blue curve shows the blueshifted modes
at high excitation power $P=150$ $\mu J/cm^{2}$; when the excitation
power is reduced, the polariton modes recover the original energy
(dashed red curve). d) Hopfield coefficients showing the exciton (XLP1,
XLP2) and photon (CLP1, CLP2) fraction of the two lower polariton
modes of (a); the green vertical line represents the in-plane momentum
of the resonantly created polariton, with LP1 having a larger exciton
fraction than LP2.}
\end{figure}

The TIR geometry chosen to perform resonant excitation allows us to
neglect the transmission through the sample, simplifying the estimation
of the incident power (see also the Supplementary Information). The
two lower polariton branches fitted by the dashed orange lines (LP1
and LP2) originate from the two optical modes highlighted by the two
black dashed lines. Polariton modes originating from different cavity
modes and coupled to the same excitonic transition possess a different
excitonic fraction at a given excitation angle. Figures 4c shows the
blueshift measured for the lower polariton branches LP1 and LP2 of
figure 4a as a function of the polariton density by using a linearly
polarized pulsed laser. We observe that, for a given incident power
(i.e. a given polariton density), the measured blueshift is higher
for the polariton branch having the higher excitonic fraction, LP1,
which confirms that the energy blueshift increases as the excitonic
component increases. By fitting the data of figure 4c with a straight
line\textbf{ }and considering the relation between\textbf{ }\textbf{\textsl{$\triangle E_{pol}$}}\textbf{
}and \textbf{\textsl{$n_{pol}$}} we can obtain the excitonic interaction
constant per layer $g_{exc,lay}$. In fact, given the incident power
values and considering that all the absorbed light is transformed
into polaritons (which is a very conservative approximation, see Supporting
Information) the upper density limit for the exciton density per layer
is of about $n_{exc,layer}=10^{12}$ $cm^{-2}$. Accounting for the
different excitonic fraction of the two lower polariton branches,
we obtain a value of the excitonic interaction constant $g_{exc,lay}\sim1\pm0.2$
$\mu eV\mu m^{2}$ per inorganic layer. This value is comparable with
that obtained for the 2D single crystal embedded in a microcavity,
as shown in Figs. 1 and 2. In both cases, we used a pulsed resonant
laser to estimate the interaction constants, thus avoiding the effects
of the polariton-reservoir interactions which would contribute to
the blueshift observed and would forbid a direct probe of the polariton-polariton
interaction constants. Moreover, we would like to stress that by neglecting
any loss in the material, we overestimate the number of excitons in
the system, therefore much larger interaction strengths should be
expected and the reported value of $g_{exc}$ should be taken as a
lower bound. Nevertheless, this lower bound is already at least two
orders of magnitude higher than typical interaction constants of organic
excitons at room temperature \citep{Daskalakis2014} and about twenty
times larger than the values recently measured in a $WS_{2}$ monolayer
\citep{Barachati2018}. 

In summary, we have observed highly interacting polaritons in hybrid
organic-inorganic 2D perovskites single crystals at room-temperature.
These materials spontaneously crystallize in a multiple layered quantum
well-like structure and, thanks to the high oscillator strength of
the excitonic transition, strong coupling is achieved at room temperature
even without highly reflecting mirrors. The resulting polaritons are
highly interacting with an excitonic interaction constant $g_{exc,lay}\geqslant1$$\pm$$0.2$
$\mu eV$$\mu m^{2}$. This value is two orders of magnitudes higher
than the values measured for organic excitons and it is the highest
measured at room temperature so far. Moreover, we observe that polariton-polariton
interactions in our sample are spin-dependent, with repulsive interaction
between polaritons having the same spin being the dominant effect.
These results demonstrate that 2D hybrid organic-inorganic perovskites
sustain highly interacting polaritons working with interaction constants
which are very promising for future polaritonic and spintronic devices
at room temperature. Most important, we observe highly interacting
polaritons at RT without embedding the active medium in an optical
cavity. These findings are extremely important and promise to greatly
reduce the cost and the complexity of fabrication and post-processing
of polaritonic devices.

\section{Methods}

\textbf{Synthesis of 2D perovskite flakes.}PEAI solutions were prepared
dissolving in gammabutyrolactone equimolar amount of PbI2 and phenethylam-
monium iodide, and stirring at 70$^\circ$C for 1 hour. 2D perovskite single
crystals were synthesized by Anti-solvent Vapor assisted Crystallization
method as follows: 5 microliters of the perovskite solution are deposited
on glass substrate and immediately after capped by a second glass.
Then, a small vial containing 2 mL of dichloromethane is placed on
the top of the two sandwiched substrates. Substrates and vial are
placed in a bigger Teflon vial, closed with a screw cap and left undisturbed
for some hours. After this time millimetre-sized crystals appear in
between the two substrates having a thickness varying from few to
ten micrometres. Single crystals are mechanically exfoliated with
SPV 224PR-M Nitto Tape and transferred onto glass substrates. The
exfoliated flakes, having the thickness of tens of nanometres, appear
smooth and uniform over tens of square micrometers, as observed by
scanning electron microscopy (SEM) and atomic force microscopy (AFM).
\begin{acknowledgments}
The authors acknowledge the ERC project ElecOpteR grant number 780757
and the project \textquotedblleft TECNOMED - Tecnopolo di Nanotecnologia
e Fotonica per la Medicina di Precisione\textquotedblright , (Ministry
of University and Scientific Research (MIUR)Decreto Direttoriale n.
3449 del 4/12/2017, CUP B83B17000010001). G.G. gratefully acknowledges
the project PERSEO-PERrovskite-based Solar cells: towards high Efficiency
and lOng-term stability (Bando PRIN 2015-Italian Ministry of University
and Scientific Research (MIUR) Decreto Direttoriale 4 novembre 2015
n. 2488, project number 20155LECAJ). The authors acknowledge Paolo
Cazzato for technical support. 
\end{acknowledgments}


\end{document}